\documentclass[aps,pra,epsfigure,twocolumn,showpacs]{revtex4}
\usepackage{color}
\usepackage{dcolumn}    
\usepackage{bm} 
\usepackage{graphicx}
\usepackage{amsmath}    
\usepackage{latexsym}
\usepackage{amsfonts}   
\usepackage{amssymb}
\usepackage{array}      
\usepackage{epsfig}
\usepackage{txfonts}

\begin{document}
\title{Criticality, factorization and long-range correlations in the anisotropic $XY$-model}
\author{Steve Campbell$^1$, Jonathan Richens$^{1,2}$, Nicola Lo Gullo$^{1,3,4}$, and Thomas Busch$^{1,4}$}
\affiliation{$^1$Quantum Systems Unit, OIST Graduate University, Okinawa 904-0495, Japan}
\affiliation{$^2$Controlled Quantum Dynamics Theory Group, Department of Physics, Imperial College London, London SW7 2AZ}
\affiliation{$^3$ Dipartimento di Fisica e Astronomia, Univestit\`a degli Studi di Padova, Padova, Italy}
\affiliation{$^4$Department of Physics, University College Cork, Republic of Ireland}

\begin{abstract}
We study the long-range quantum correlations in the anisotropic $XY$-model. By first examining the thermodynamic limit we show that employing the quantum discord as a figure of merit allows one to capture the main features of the model at zero temperature. Further, by considering suitably large site separations we find that these correlations obey a simple scaling behavior for finite temperatures, allowing for efficient estimation of the critical point. We also address ground-state factorization of this model by explicitly considering finite size systems, showing its relation to the energy spectrum and explaining the persistence of the phenomenon at finite temperatures. Finally, we compute the fidelity between finite and infinite systems in order to show that remarkably small system sizes can closely approximate the thermodynamic limit.
\end{abstract}
\date{\today}
\maketitle

\section{Introduction}
\label{Intro}
The study of many-body systems is a very active area of research, motivated by an acute observation by Anderson over forty years ago~\cite{anderson}: {\it more is different}. A system made of many bodies is not simply the sum of them, but something more complicated. In other words, we cannot expect that the behavior 
of a many-body system is understood once the physics of its constituent parts is known. Interactions, no matter how weak, significantly enrich the range of observable phenomena. Due to these interactions many-body systems can appear in different phases each of them with peculiar properties. In the case of quantum systems we have quantum phase transitions (QPTs), which occur at zero temperature where thermal fluctuations are absent. In fact, they are driven by quantum fluctuations, which are fluctuations in the mean value of observables of a system due to the Heisenberg uncertainty principle. Most of the known QPTs are well described in the Ginzburg-Landau picture where the change from one phase to another is accompanied by a symmetry breaking process and the consequent development of a non-zero value for some order parameter.

One of the most striking facts about QPTs is their universality~\cite{sachdev}. This means that different systems exhibit the same behavior at the critical (or transition) point, regardless of the microscopic details, {\it e.g.} the type of interaction or nature of the system. Therefore, and without loss of generality, we choose here to study exactly solvable spin-models in order to further understand QPTs, and more generally criticality in quantum systems. In this regard, the $XY$-model holds particular appeal because in addition to a QPT it also possesses another peculiar phenomenon: {\it factorization}~\cite{kurmann,adesso}. Spin systems in an external magnetic field can show a fully factorized state in the ordered phase, {\it i.e.} the phase in which spin-spin interactions prevails over the external field and the system is free to self-organize. Early explanations involved the analysis of pairwise entanglement around the factorization field, $\lambda_f$. These studies showed that across $\lambda_f$ the two spin entanglement undergoes a change from parallel to anti-parallel~\cite{fubini,palma}, being zero exactly at $\lambda_f$. For this reason ground-state factorization has been referred to as an ``entanglement transition". Recently it has been related to a  change in the symmetry of the ground state~\cite{rossignoli1,gianluca} indicating its fundamental importance.

Exploring both criticality and factorization using the tools of quantum information has proven fruitful~\cite{kurmann,adesso,fubini,fubini2,palma,rossignoli1,gianluca,amico1,amico2,amico3,osborne,fazio,dillenschneider,sarandy1,maziero1,sarandy2,maziero2,turks,huang,campbellGD,gabriele1,gabriele2}. While most studies consider only nearest neighbor pairs of spins, exploiting favourable figures of merit allows to access longer ranges~\cite{maziero2,turks}, finite temperatures~\cite{turks,campbellGD}, and  finite sizes~\cite{rossignoli1,campbellGD,gabriele2}. Small finite size systems also allow for the study of multipartite correlations~\cite{campbellGD,monogamy,rulli,campbellNL}, an important topic in itself. Here we show that a general figure of merit for quantum correlations, namely the quantum discord, is a versatile for tool to studying criticality and factorization, particularly in situations where entanglement is either severely constrained or has become identically zero. By studying long-range pairs in the thermodynamic limit we find that the quantum discord captures the main features of criticality and obeys a simple function for critical point estimation at finite temperatures. Furthermore we find that the qualitative features of factorization, both at zero and finite temperature, can be explained by studying small systems and that such systems closely approximate the thermodynamic limit.

The remainder of the paper is organized as follows. In Sec.~\ref{toolbox} we introduce the anisotropic $XY$-model and tools used throughout the paper. Then, in Sec.~\ref{thermodynamiclimit} we study pairs of spins in the thermodynamic limit, and show the versatility of long-range correlations for studying criticality at both zero and finite temperature. We address the factorization phenomenon by studying finite size systems in Sec.~\ref{finitesize} . The fidelity between the finite size states and the exactly solved thermodynamic limit is calculated in Sec.~\ref{fidelity},  and in Sec.~\ref{conclusions} we conclude.

\section{Preliminaries}
\label{toolbox}
Let us first introduce the system under study and the 
mathematical treatment that allows us to calculate the quantities
which will be the focus of our discussion in the remainder of the paper.
We recall the definitions of the two main figures of merit, namely
the entanglement of formation and the quantum discord.

\subsection{The Model}
We consider the anisotropic $XY$-model with periodic boundary conditions 
and assuming only nearest-neighbor interaction the Hamiltonian is given by
\begin{equation}
\label{XY}
\begin{aligned}
\mathcal{H}=-\sum_{i=0}^{N-1}\left[\frac{\lambda}{2} \left( [1+\gamma] \sigma_x^{i} \otimes \sigma_x^{i+1}+[1-\gamma] \sigma_y^{i} \otimes \sigma_y^{i+1} \right) + \sigma_z^i \right],
\end{aligned}
\end{equation}
where $\lambda$ is the spin-spin interaction strength, $\gamma\!\in\![0,1]$ is the anisotropy parameter, and $\sigma_{x,y,z}$ are the usual Pauli operators. For the forthcoming discussions one should note that the above Hamiltonian
is invariant under parity transformation: $[\mathcal{H},P]=0$ with 
$P=e^{i\frac{\pi}{2}(\sum_i \sigma_z^{i}+N)}$~\cite{rossignoli1}.
This implies that any {\it non-degenerate} eigenstate of the Hamiltonian, and in particular its ground
state, is also an eigenstate of the parity operator.
A quantity that captures important aspect of the behavior of the model is the so-called two spin reduced density matrix, which is readily obtained in the thermodynamic limit, $N\!\to\!\infty$, by expressing it in terms of the two-point correlation functions and the magnetization~\cite{mccoy}. For two spins in the chain separated by $r$ sites it is given by 
 \begin{equation}
 \label{densitymatrix}
 \varrho_{0r}=\frac{1}{4}\left(  \openone +\big<\sigma_z\big> (\sigma_z^0+\sigma_z^r) +\sum_{i=x,y,z} \big< \sigma_i^0 \sigma_i^r \big> \sigma_i^0 \sigma_i^r   \right),
 \end{equation}
where the two point correlation functions are defined as
\begin{eqnarray}
&\big< \sigma_x^0 \sigma_x^r \big>&= 
\begin{vmatrix}
  G_{-1} & G_{-2} & \cdots & G_{-r} \\
  G_{0} & G_{-1} & \cdots & G_{-r+1} \\
  \vdots  & \vdots  & \ddots & \vdots  \\
  G_{r-2} & G_{r-3} & \cdots & G_{-1}
 \end{vmatrix},\\
&\big< \sigma_y^0 \sigma_y^r \big>&= 
\begin{vmatrix}
  G_{1} & G_{0} & \cdots & G_{-r+2} \\
  G_{2} & G_{1} & \cdots & G_{-r+3} \\
  \vdots  & \vdots  & \ddots & \vdots  \\
  G_{r} & G_{r-1} & \cdots & G_{1}
 \end{vmatrix},\\
&\big< \sigma_z^0 \sigma_z^r \big>&= \big<\sigma_z\big>^2-G_rG_{-r}.
\end{eqnarray}
The function $G_r$, the magnetization $\big< \sigma_z \big>$, and $\omega_\phi$, are given by
\begin{equation}
\begin{aligned}
&G_r= \int^\pi_0 d\phi \frac{\tanh(\beta\omega_\phi)}{2\pi\omega_\phi} [ \cos(r\phi)(1+\lambda\cos\phi)- \\
         &\hskip3.5cm\lambda~\gamma\sin(r\phi)\sin\phi ],
\end{aligned}
\end{equation}
\begin{eqnarray}
&\big< \sigma_z \big>&=-\int^\pi_0 d\phi \frac{(1+\lambda\cos\phi)\tanh(\beta\omega_\phi)}{2\pi\omega_\phi},\\
&\omega_\phi&=\frac{1}{2}\sqrt{(\lambda~\gamma\sin{\phi})^2 + (1+\lambda\cos\phi)^2},
\end{eqnarray} 
and $\beta=1/T$ is the inverse temperature.

While the literature about this system is already quite extensive, most studies focus on the QPT  at $T=0$ by studying nearest- or next-nearest-neighbor correlations~\cite{osborne,fazio,dillenschneider,sarandy1,maziero1} and only recently have longer ranges been considered~\cite{maziero2,turks}. Here we rigorously assess the differences and advantages arising from studying long-range ground-state and thermal quantum correlations in understanding criticality and factorization.
The latter describes the existence of a value for the external field at which the ground state of the system at zero temperature becomes fully factorized. For the Hamiltonian given by Eq.~(\ref{XY}), this factorization field is given by
\begin{equation}
\label{factor}
\lambda_f=\frac{1}{\sqrt{1-\gamma^2}}.
\end{equation}

\subsection{Figures of merit for quantum correlations}
Our discussion will focus on the differences in the behavior of two figures of
merit, the entanglement of formation (EoF) and quantum discord (QD). Due to their construction they share the same entropic definition and, for pure states, they both are equivalent to the von Neumann entropy. Studying the distribution of quantum correlations in multipartite states allows for a relationship connecting bipartite QD and EoF to be established~\cite{fanchini2}, therefore making them the most natural choices for qualitative and quantitative comparison. This relationship has recently been examined in~\cite{fanchini1}.

QD can be expressed as the difference between two classically equivalent versions of mutual information, that measure the total correlations within a quantum state~\cite{discord1,discord2,paternostro}. For a two-qubit state $\rho_{AB}$, the mutual information is 
\begin{equation}
{\cal I}(\rho_{AB})={\cal S}(\rho_A)+{\cal S}(\rho_B)-{\cal S}(\rho_{AB}), 
\end{equation}
where, ${\cal S}(\rho)\,{=}\,{-}\text{Tr}[\rho\log_2\rho]$ is the von Neumann entropy of a generic state $\rho$. One can also define the one-way classical correlations~\cite{discord2} 
\begin{equation}
{\cal J}^\leftarrow(\rho_{AB})={\cal S}(\rho_{A})-{\cal H}_{\{\Pi_i\}}(A|B), 
\end{equation}
where we have introduced ${\cal H}_{\{\Pi_i\}}(A|B){\equiv}\sum_{i}p_i{\cal S}(\rho^i_{A|B})$ as the quantum conditional entropy associated with the the post-measurement density matrix $\rho^i_{A|B}=\text{Tr}_{B}[\Pi_i\rho_{AB}]/p_i$ obtained by performing a complete projective measurement $\{\Pi_i\}$ on qubit $B$. This leads to the QD to be defined as
\begin{equation}
{\cal D}^\leftarrow\,{=}\,\inf_{\{\Pi_i\}}[{\cal I}(\rho_{AB})-{\cal J}^\leftarrow(\rho_{AB})],
\end{equation}
with the infimum calculated over the set of projectors $\{\Pi_i\}$ \cite{discord1,paternostro}. ${\cal D}^\rightarrow$ is obtained simply by swapping the roles of $A$ and $B$. 
Since the states given by Eq.~(\ref{densitymatrix}) are symmetric they do not suffer the asymmetry usually associated with the QD and we will simply refer to $\mathcal{D}$ as the QD of the state regardless of which subsystem is measured.

EoF is an entanglement monotone that quantifies the minimum number of Bell pairs needed in order to prepare a copy of the state in question~\cite{wootters}. For any two-qubit state the EoF is calculated as
\begin{equation}
\mathcal{E}\,{=}\,h\left(\frac{1}{2}\left[1+\sqrt{1-\mathcal{C}^2}\right]\right),
\end{equation}
where $h(x)\,{=}\,-x\text{log}_2 x - (1-x)\text{log}_2 (1-x)$ is the binary entropy function and $\mathcal{C}$ is the concurrence of the state~\cite{wootters}. The latter is an equally valid entanglement measure and can be found in terms of the eigenvalues $\lambda_1\,{\geq}\, \lambda_{2,3,4}$ of the spin-flipped density matrix $\rho_{AB}(\sigma_y\otimes\sigma_y)\rho^*_{AB} (\sigma_y\otimes\sigma_y)$ as
\begin{equation}
\mathcal{C}=\text{max}\left[0,\sqrt{\lambda_1}-\sum_{i=2}^{4}\sqrt{\lambda_i}\right].
\end{equation}

\section{Long-range correlations in the thermodynamic limit}
\label{thermodynamiclimit}
In this section we describe the behavior of the above figures of merit for quantum correlations in the thermodynamic limit, $N\!\to\!\infty$. We will address the cases of the ground state, $T\!\to\!0$, and thermal state, $T>0$, separately in order to compare our results with existing literature more easily. It is important to notice that when discussing the ground state we will refer to the thermal ground state of the system.

\subsection{Ground state case (T$\to$0)}
\begin{figure}[b]
{\bf (a)}\hskip3.5cm{\bf (b)}\\
\includegraphics[scale=0.31]{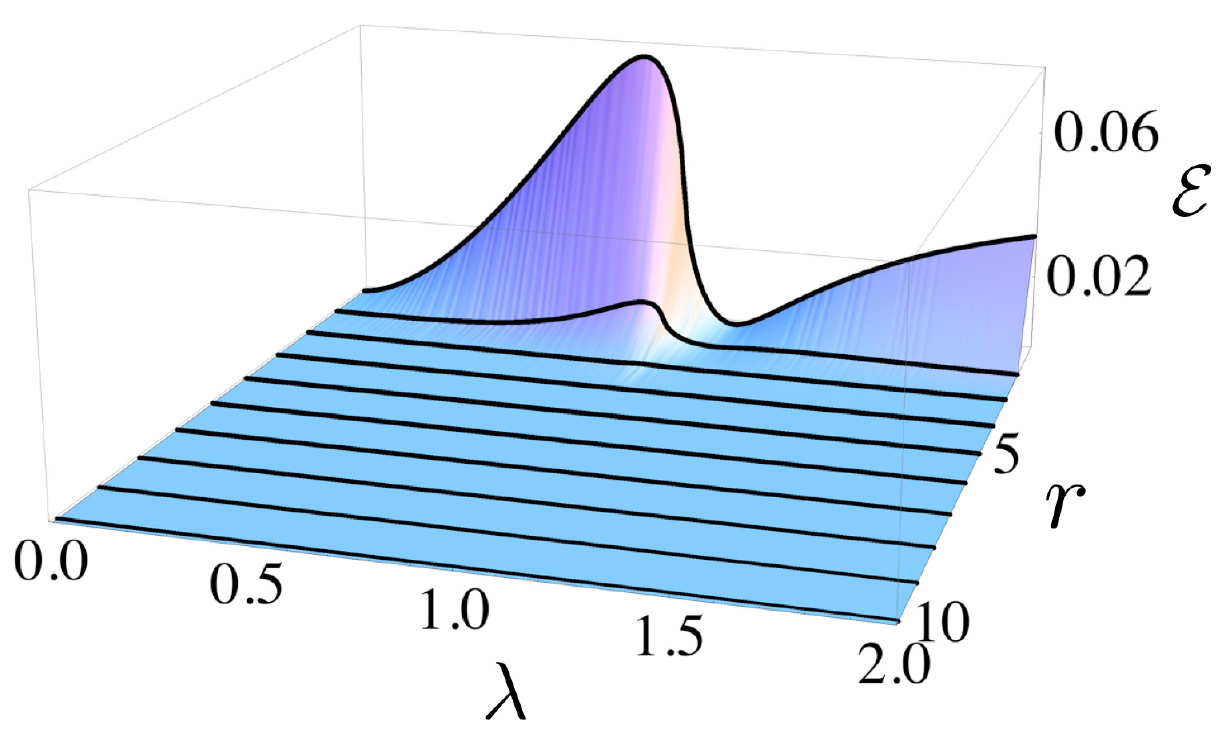}~~~\includegraphics[scale=0.31]{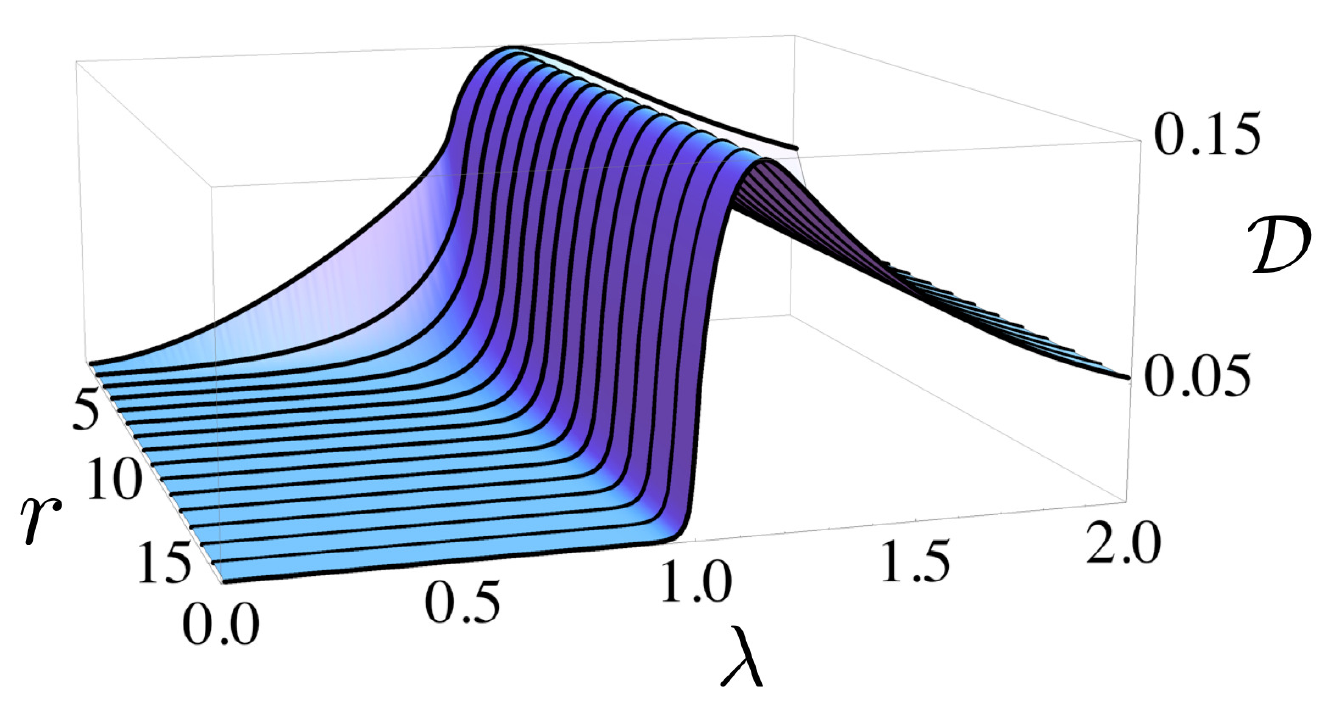}
\caption{{\bf (a)} EoF and {\bf (b)} QD as a function of coupling strength, $\lambda$, and site-separation, $r$, for fixed anisotropy, $\gamma=0.5$. While entanglement quickly decays the QD has a non-trivial behavior at long-ranges. Note that the smooth underlying curve is just a guide to the eye.}
\label{gamma}
\end{figure}
In Fig.~\ref{gamma} we show the behavior of the EoF and QD for a fixed value of the anisotropy, $\gamma=0.5$. In panel {\bf (a)} one can see that the pairwise entanglement decays quickly for increasing separation and for $r>2$ it is almost identically zero. As discussed in~\cite{osborne} this can be understood due to the constraints on the sharing of bipartite entanglement, which must scale inversely with $N$. The finite range of entanglement around the factorization field is analyzed in~\cite{fubini2}. In contrast, panel {\bf (b)} shows that the QD exhibits a much richer behavior, displaying an equally complex behavior for short ranges, while becoming more uniform with increasing $r$. However, as QD is not constrained in the same manner as entanglement we see that it can maintain quite large non-zero values for any $r$. While it is well established that for short ranges both figures of merit capture the QPT~\cite{osborne,fazio,sarandy1}, it is interesting that only the QD appears to capture the main features of the two phases for all $r$. In the ferromagnetic phase, $\lambda > 1$, QD is larger than in the paramagnetic phase, $\lambda < 1$ and a sharp change at the critical point, $\lambda_c=1$, is visible. Indeed the long-range QD embodies the QPT mechanism, as understood in the Gizburg-Landau picture~\cite{amico1}, shown in panel {\bf (b)}: it approaches zero in the paramagnetic phase but it has a finite jump across the critical point as the system enters the ferromagnetic phase.

\begin{figure}[t]
{\bf (a)}\hskip3.5cm{\bf (b)}\\
\includegraphics[scale=0.37]{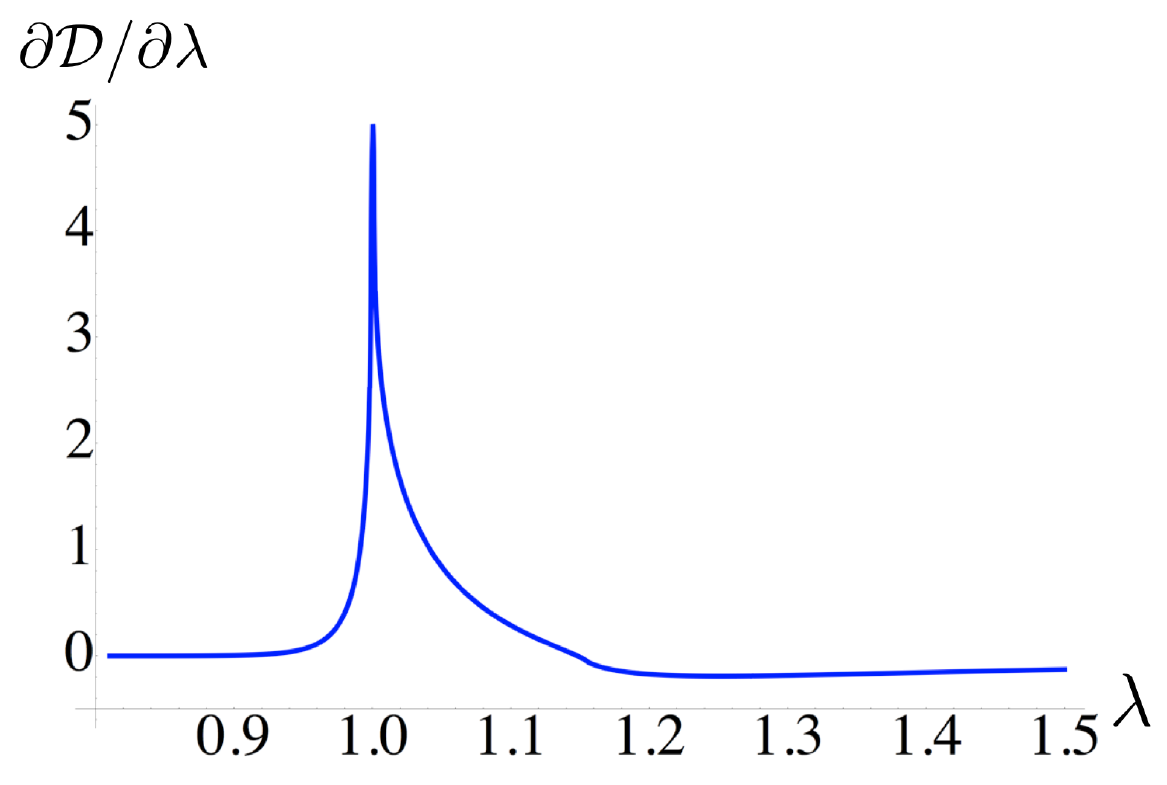}~~~\includegraphics[scale=0.4]{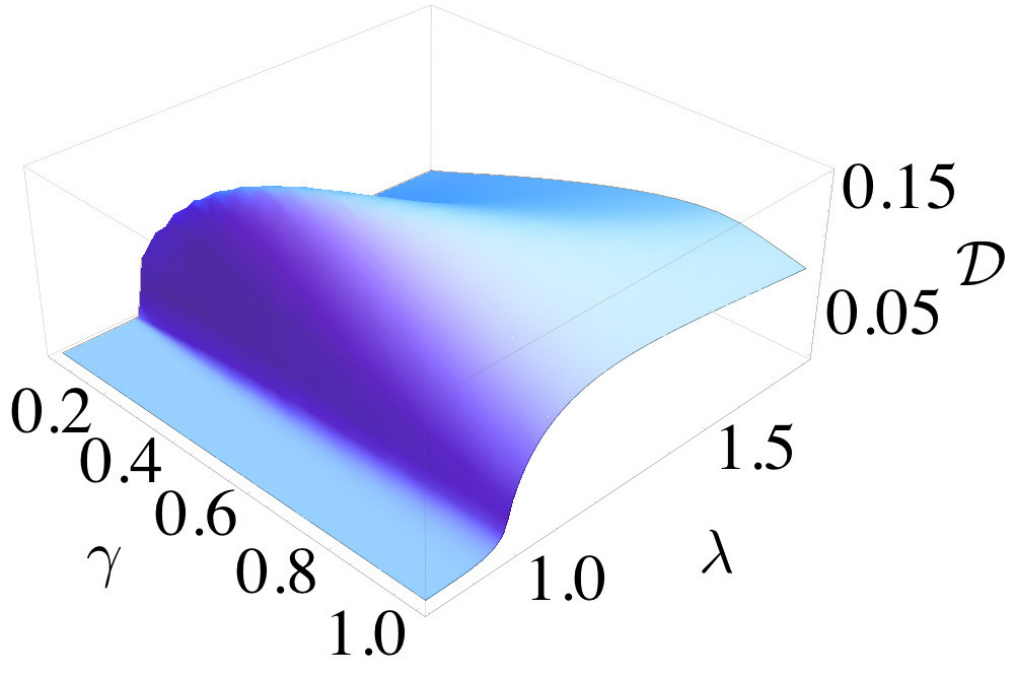}
\caption{{\bf (a)} Behavior of the first derivative of QD with respect to $\lambda$, $\partial\mathcal{D}/\partial\lambda$, as a function of $\lambda$ for $\gamma=0.5$ and $r=15$. {\bf (b)} Behavior of QD against anisotropy $\gamma$ and coupling $\lambda$ of the ground state, $T=0$, for site separation $r=15$.} 
\label{DiscordN15}
\end{figure}

Recall that the critical behavior of the system is universal, {\it i.e.} it does not depend on the microscopic details, in particular on the nature of the short range interaction. This universality is captured quite strikingly by the long-range QD, and together with the scaling behavior of the QD shown in~\cite{amico1,huang} suggests that, while both figures of merit faithfully capture the QPT, QD presents a much clearer behavior. When examining nearest- and next-nearest-neighbors, both QD and EoF exhibit a discontinuity in their derivative with respect to $\lambda$ as we move across the critical point. Interestingly, the QD maintains the same strong discontinuity regardless of $r$, shown in Fig.~\ref{DiscordN15} {\bf (a)}, where we plot the derivative of QD with respect to $\lambda$ when $\gamma=0.5$ and for a site separation of $r=15$. The sudden onset of long-range correlations is one characteristic of a QPT~\cite{sachdev}. Only in regions tightly confined around $\lambda=\lambda_c$ does entanglement allow for such a feature to be witnessed, and here the maximum separation between spins showing a non-zero entanglement scales with $\gamma^{-1}$~\cite{fazio}. However, again due to the constraints on its shareability the actual value of entanglement decreases exponentially with increasing distance, and outside of this tight region around $\lambda=1$ it is zero. In contrast we find this onset of long-range correlations in the QD for all values of anisotropy, shown in Fig.~\ref{DiscordN15} {\bf (b)}, and all values of $r$~\cite{huang}.

\begin{figure}[b]
{\bf (a)}\hskip3.5cm{\bf (b)}\\
\includegraphics[scale=0.4]{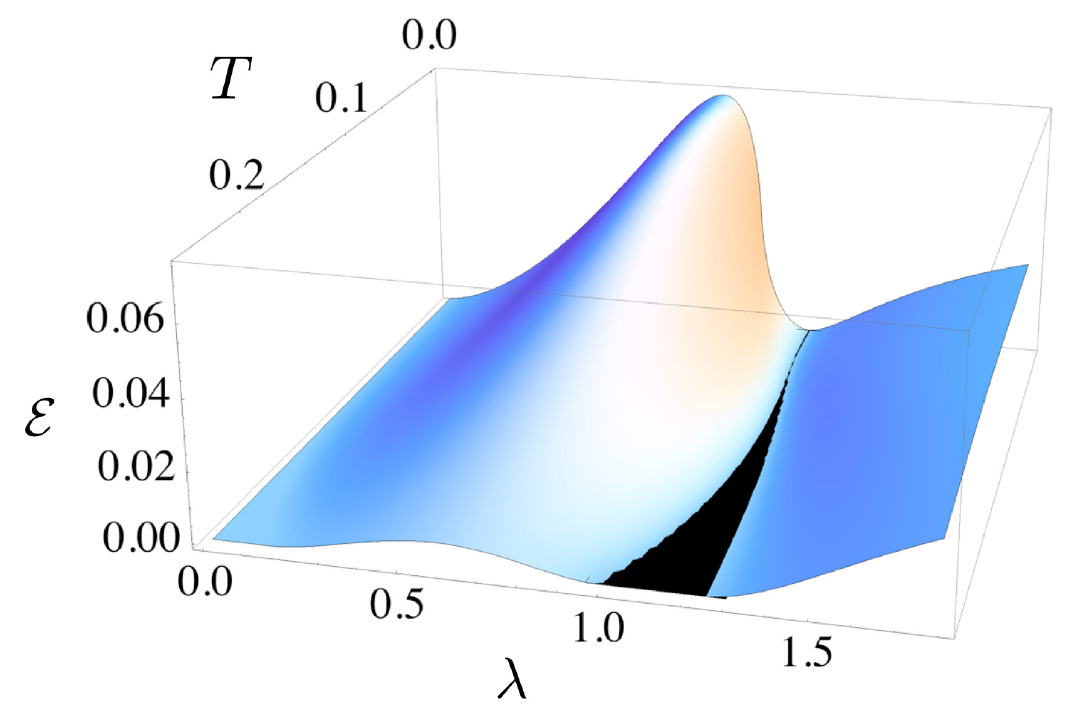}~~\includegraphics[scale=0.4]{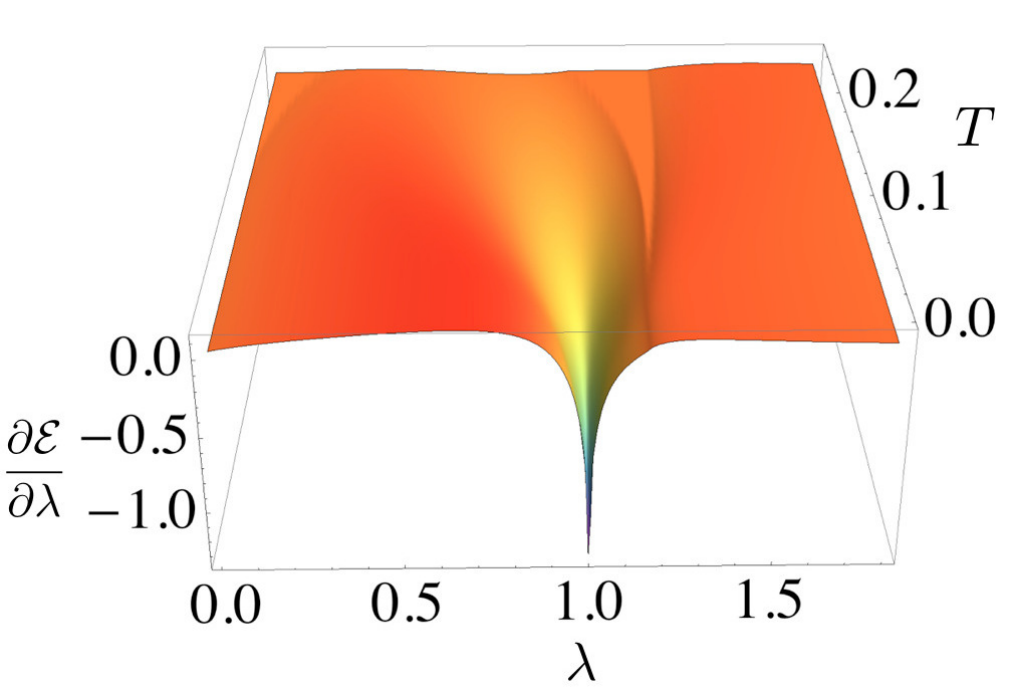}\\
{\bf (c)}\hskip3.5cm{\bf (d)}\\
\includegraphics[scale=0.4]{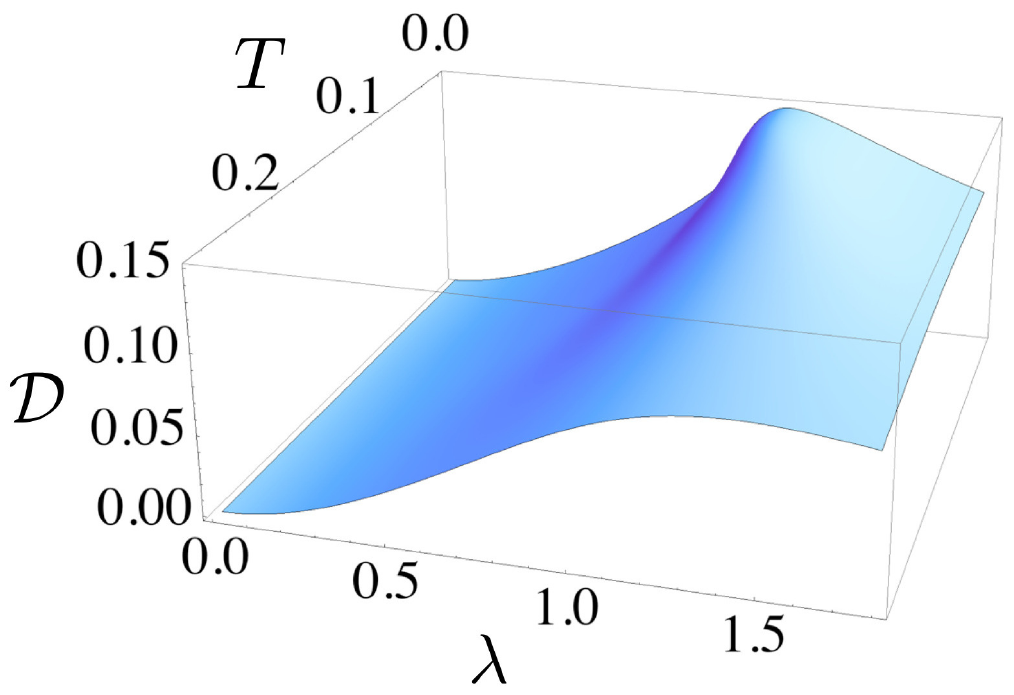}~~\includegraphics[scale=0.4]{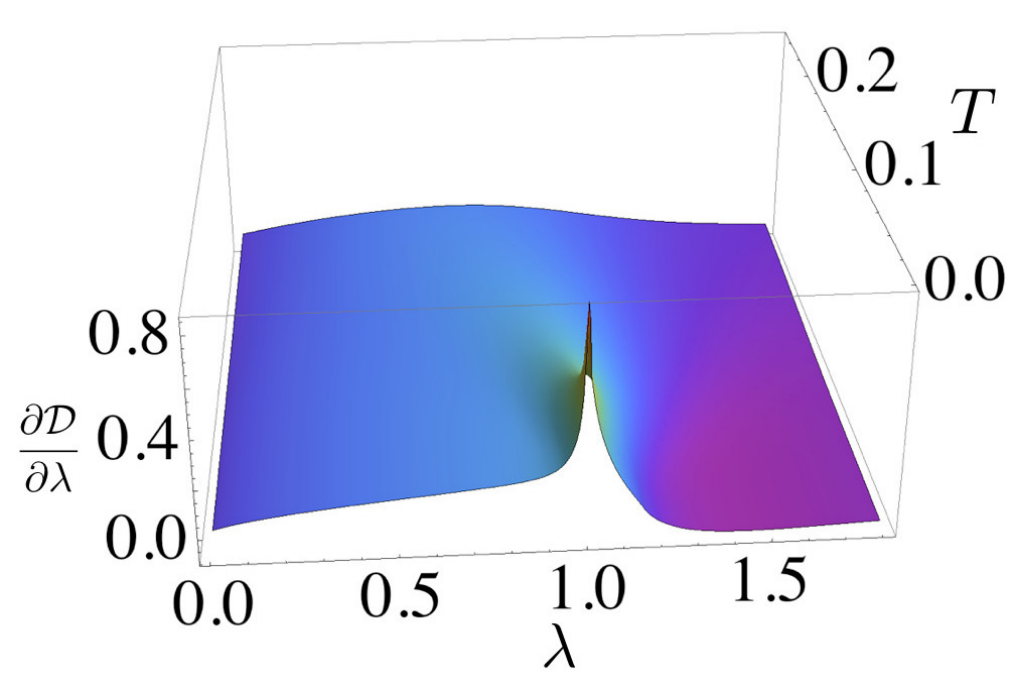}\\
{\bf (e)}\hskip3.5cm{\bf (f)}\\
\includegraphics[scale=0.4]{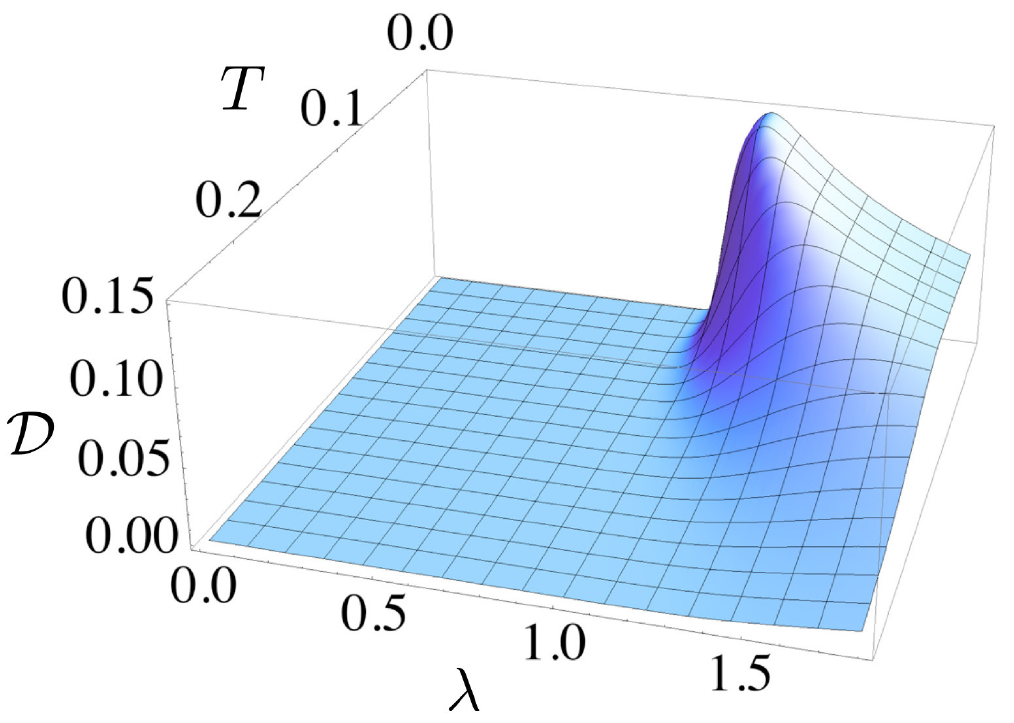}~~\includegraphics[scale=0.4]{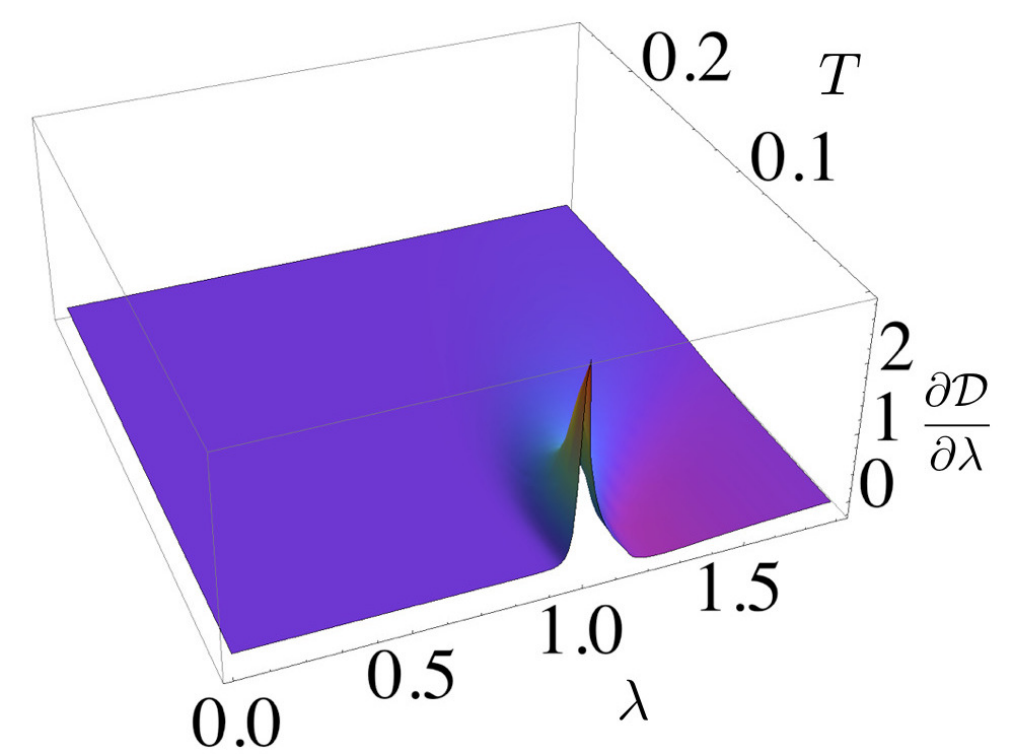}\\
\caption{Behavior of correlations at finite temperature for $\gamma=0.5$. {\bf (a)} Nearest-neighbor EoF. The black plane is $\mathcal{E}=0$, and as $T$ increases the spreading out of the factorization point into a region of separability can be seen. {\bf (b)} Derivative of nearest neighbor EoF,~~$\partial\mathcal{E}/\partial\lambda$,~ as a function of $\lambda$ and $T$. {\bf (c)} Nearest-neighbor QD. {\bf (d)} Derivative of nearest neighbor QD,~~$\partial\mathcal{D}/\partial\lambda$,~ as a function of $\lambda$ and $T$. {\bf (e)} Long-range QD for $r=15$ and {\bf (f)} its derivative $\partial\mathcal{D}/\partial\lambda$ as a function of $\lambda$ for $r=15$. Notice in all plots of the derivatives, $T$ goes into the page.}
\label{finiteT}
\end{figure}

\subsection{Thermal case (T$>$0)}
The behavior of critical spin systems at finite temperature has been an active area of research recently~\cite{maziero2,turks,campbellGD}. While typically detrimental, considering the effects of finite temperature is extremely important, both when trying to understand the nature of criticality and the limitations of any realistic experimental attempts to witness such phenomena. Strictly speaking, a QPT is defined only at $T=0$ and relaxing this constraint means looking for signatures of the critical nature in situations where the characteristic behaviors have been degraded by the mixing of higher energy levels. This usually leads to the critical point becoming a critical region~\cite{sachdev,amico1}. In Fig.~\ref{finiteT} {\bf (a)} and {\bf (c)} we show the behavior of the nearest-neighbor EoF and QD respectively. One can immediately see that the entanglement decays with increasing $T$ and, more interestingly, that the factorization point spreads out (black area). For QD we see that finite $T$ smooths out and gradually decreases the quantum correlations present between nearest neighbors. These effects can also be seen in the behavior of the derivatives of the EoF [$\partial\mathcal{E}/\partial\lambda$ panel {\bf (b)}] and the QD [$\partial\mathcal{D}/\partial\lambda$ panel {\bf (d)}]. For both quantities a sharp discontinuity at $\lambda=\lambda_c$ quickly smooths out for finite temperatures. As has been proposed before, using the behavior of the derivative to estimate the critical point by identifying its extremal points for non-zero $T$ is a reasonable approach~\cite{amico1,turks}, and we refer to this as the estimated thermal critical point (ETCP), $\lambda_{T_c}$. For nearest- and next-nearest-neighbors the competition between the interaction and thermal effects cause the ETCP to deviate from the critical point for finite $T$, with the actual deviation varying significantly depending on the separation and correlation measure employed~\cite{turks}.

Long-range correlations are again only captured by the QD and in  panels {\bf (e)} and {\bf (f)} of Fig.~\ref{finiteT} we show the QD and its derivative as a function of $\lambda$ and $T$ for a site separation of $r=15$. Once again one can see a sharp increase in the QD near the critical point and the associated discontinuity in its derivative is smoothed out as we reach higher $T$. However dealing with long-range correlations has one subtle but crucial aspect in that we now see the behavior become more uniform. The ETCP increasingly shifts away from $\lambda_c$ as $T$ increases which is due to the fact that long-range correlations effectively ignore features that arise due to the short range nature of the interaction and focus on the global properties of the system. 

\begin{figure}[t]
{\bf (a)}\\
\includegraphics[scale=0.7]{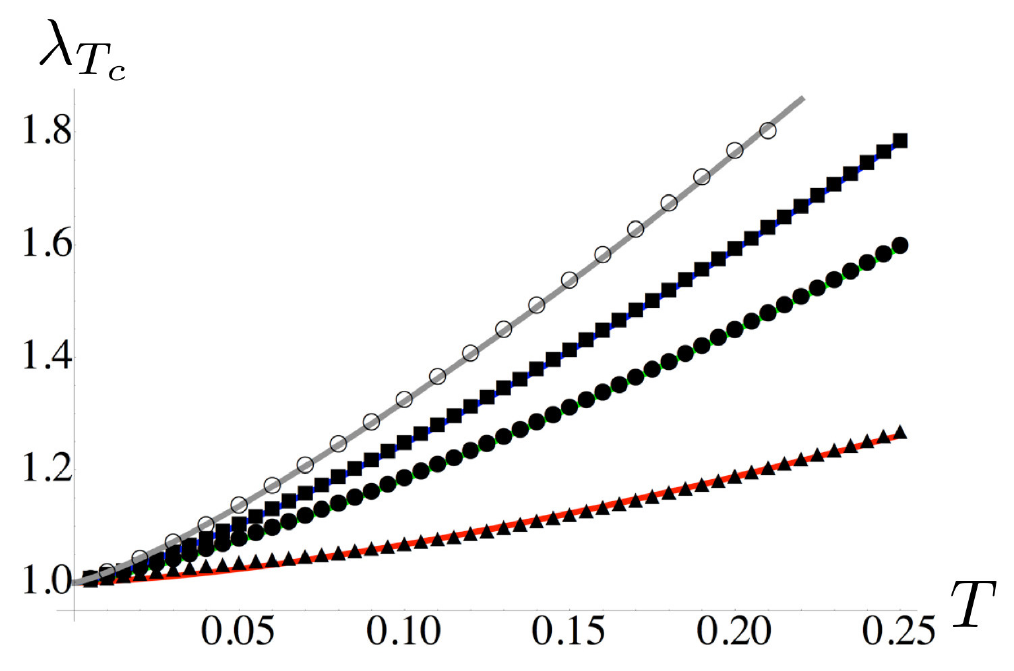}\\
{\bf (b)}\\
\includegraphics[scale=0.7]{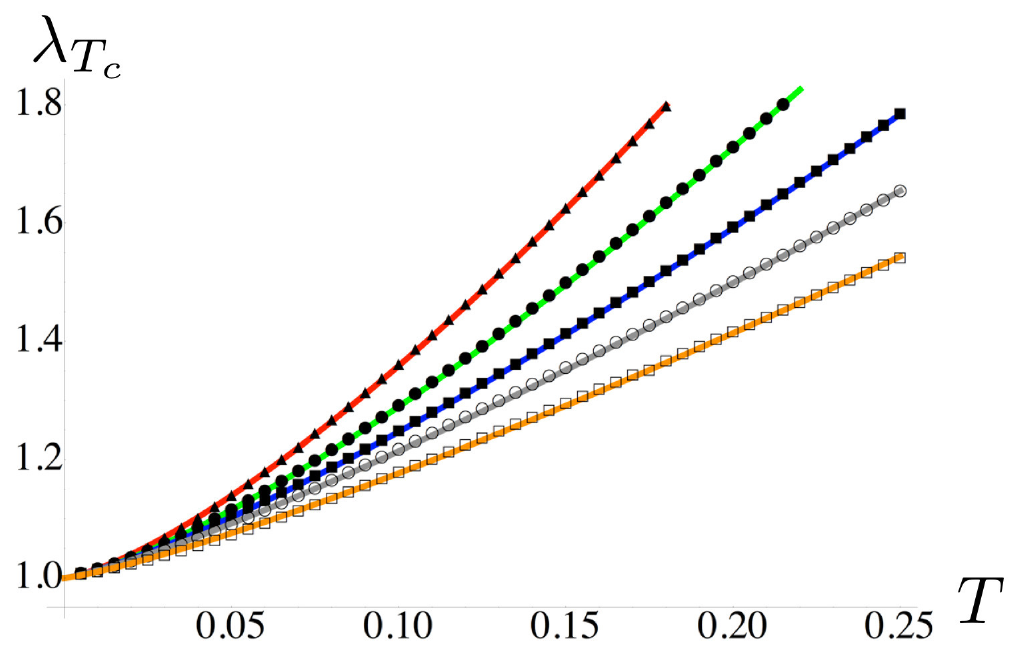}
\caption{Estimated thermal critical point, $\lambda_{T_c}$, determined by identifying the maximum of the derivative of the QD for a thermal state. {\bf (a)} Fixed $\gamma=0.5$ and increasing site-separations $r=5$ [triangles], 10 [filled circles], 15 [filled squares], and 25 [empty circles]. {\bf (b)} Fixed $r=15$ for various values of anisotropy $\gamma=0.15$ [triangles], 0.3 [filled circles], 0.5 [filled squares], 0.75 [empty circles], and 1 [empty squares]. The curves drawn through each point set in both panels are the best fit satisfying the ansatz, Eq.~(\ref{universalT}).}
\label{criticalpointest}
\end{figure}

To capture this uniform behavior for $\lambda_{T_c}$ we suggest the ansatz
\begin{equation}
\lambda_{T_c}=\alpha~{T}^\nu+1,
\label{universalT}
\end{equation}
and in Fig.~\ref{criticalpointest} {\bf (a)} compare it to the numerically obtained values for different separations $r$ = 5, 10, 15, and 25. For fixed $\gamma=0.5$, the point markers are the ETCPs for various values of $T$ and the lines correspond to a best fit of for the parameters $\alpha$ and $\nu$ (see Table~\ref{table1}). We see that the gradient increases significantly as we increase the site separation and for $r=5$ the exponent $\nu$ is notably different from the ones for larger separations. Such a deviation indicates that the coupling effects of neighboring spins are still present. In Fig.~\ref{criticalpointest} {\bf (b)} we show that the scaling behavior suggested in Eq.~(\ref{universalT}) is also valid when varying the anisotropy parameter, $\gamma$. As we increase $\gamma$ the values for the gradient and exponent change more significantly, although a clear trend is still present and larger anisotropy leads to a decrease in the gradient $\alpha$. Other similar scaling behaviors are discussed in~\cite{amico1,amico3}, however the simple dependence of $\partial\mathcal{D}/\partial\lambda$ up to large $T$ for long-ranges is quite a remarkable result, indicating another clear advantage of using long-range quantum correlations to study criticality. 

Returning to the factorization field one can see that for finite $T$ this phenomena disappears. Indeed, for the ground state factorization can be witnessed in a number of ways, in particular, by examining short range ($r<3$) entanglement and finding the point when it is zero. Additionally one can calculate the QD and identify when it takes a value independent of $r$. When one examines thermal cases, these features are lost for $T>0.03$~\cite{amico1}. At this temperature the factorization point starts to spread into a region of separability, as shown by the behavior of nearest neighbor EoF in Fig.~\ref{finiteT} {\bf (a)}, in which one will not find a non-zero constant value of QD independent of $r$. This indicates that the factorization field requires a more detailed analysis.

\section{Understanding the factorization Field: finite size considerations}
\label{finitesize}
\begin{figure}[t]
{\bf (a)}\hskip3.5cm{\bf (b)}\\
\includegraphics[scale=0.4]{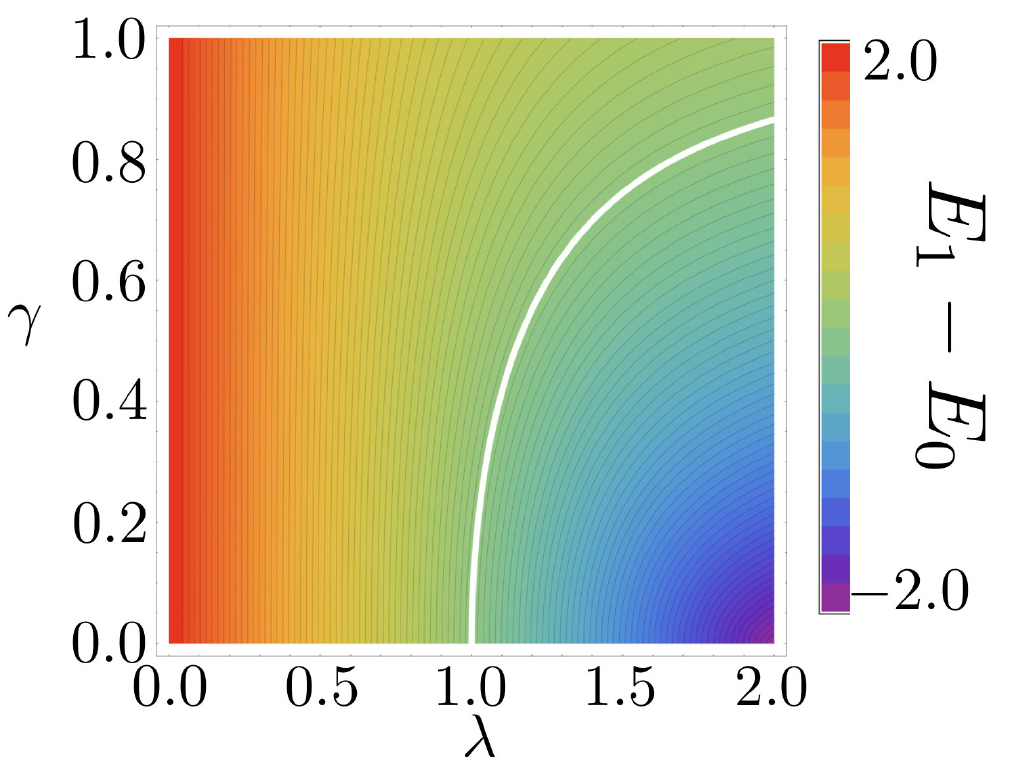}~\includegraphics[scale=0.4]{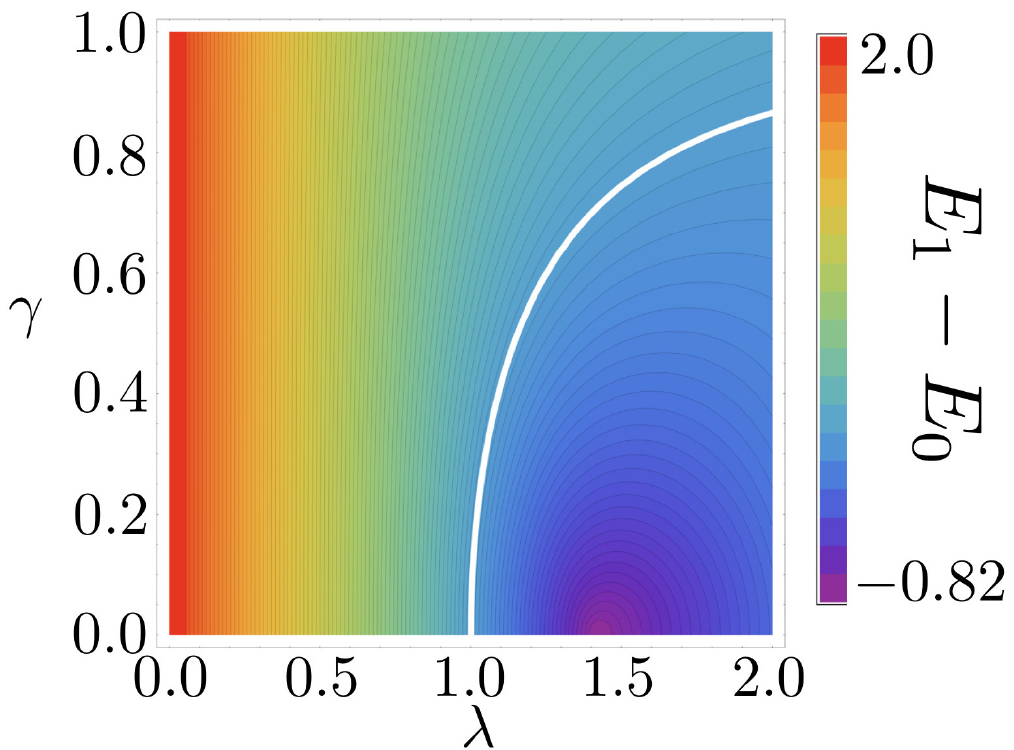}\\
{\bf (c)}\hskip3.5cm{\bf (d)}\\
\includegraphics[scale=0.4]{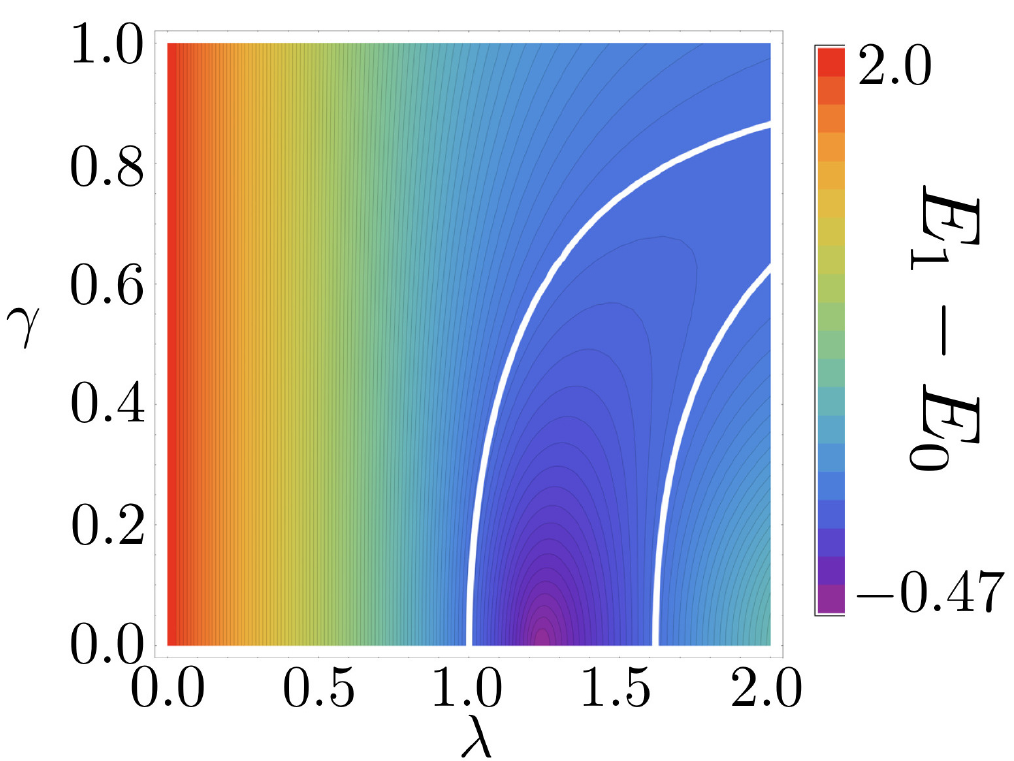}~\psfig{figure=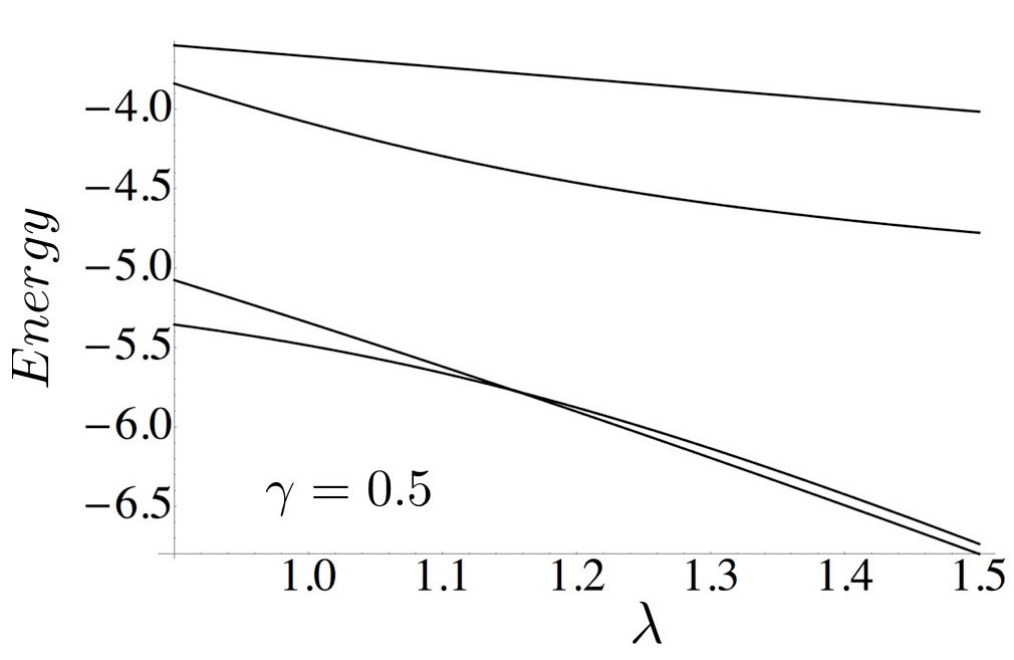,width=4cm,height=3.2cm}
\caption{Difference between first excited, $E_1$, and ground state energy, $E_0$, for finite size chains of {\bf (a)} 3 spins {\bf (b)} 4 spins and {\bf (c)} 5 spins. Regardless of the chain size the first energy level crossing, where $E_1-E_0=0$, always appears along the factorization field given by Eq.~(\ref{factor}), indicated by the leftmost white line in each panel. {\bf (d)} Lowest four energy levels for $N=5$ and $\gamma=0.5$. For clarity we restrict $\lambda\in[0.9,1.5]$.}
\label{energylevels}
\end{figure}

In this section we examine the behavior of finite sized systems at the factorization field, and show that factorization can be understood in terms of an energy level crossing. This approach has been discussed before~\cite{gianluca,rossignoli1,adesso} and here we present some additional observations as well as explicitly considering the implications of this explanation at finite $T$. As recently explored by some of us, departing from the thermodynamic limit and considering small finite sized systems still allows for the study of the interesting properties of many-body quantum systems~\cite{campbellGD}. In Fig.~\ref{energylevels} we plot the difference between the energies of the ground and first excited state for finite size chains of lengths $N=3,4$ and $5$~\cite{energynote}. One can see that regardless of the system size the model exhibits an energy level crossing at a point exactly coinciding with the factorization field, as highlighted by the white line. For the smallest non-trivial ring, $N=3$, this is the only energy level crossing, while for increasing $N$ one finds $\frac{N}{2}$ $\left(\frac{N-1}{2}\right)$ crossings for $N$ even (odd)~\cite{gianluca}. 

\begin{table}[t]
\centering
\begin{tabular}{|c|ccc||c|ccc|}
\hline 
&  & $\gamma=0.5$ & & & &$r=15$ &\\ [0.5ex] 
\hline 
$r$ & $\alpha$ & & $\nu$ & $\gamma$ & $\alpha$ & & $\nu$\\ [0.5ex]
\hline 
~~5~~&~2.01796 & & 1.47349 &~~0.15~~&~8.26481 & & 1.36232  \\
~~10~~&~3.5269 & & 1.28208 &~~0.30~~&~6.17000 & & 1.32828 \\
~~15~~&~4.50366 & & 1.26092 &~~0.75~~&~3.53507 & & 1.21671\\
~~25~~&~5.63417 & & 1.24251 &~~1.00~~&~2.96397 & & 1.22295 \\ 
\hline
\end{tabular}
\caption{Parameter values corresponding to the curves of best fit for Eq.~(\ref{universalT}) shown in Fig.~\ref{criticalpointest}.}
\label{table1}
\end{table}

By examining the ground state of such finite size systems we can now explain why there is constant value of QD for all $r$ at $\lambda_f$ in the thermodynamic limit. At the factorization point the ground state of the system is twofold degenerate and the two states are highly symmetric with opposite parity. Thus the thermal ground state is an equal mixture of these degenerate eigenstates and the reduced $M\times M$ density matrices for any choice of $M(\!<\!N)$ subsystems are identical~\cite{rossignoli1}. From this it follows straightforwardly that the correlations (EoF or QD) take a constant value regardless of what pair of spins one chooses to look at. However, this is not the case at any of the other energy level crossings when $N>3$. 

The understanding of factorization as related to an energy level crossing also explains why the phenomenon disappears when dealing with suitably large finite $T$ and sheds light on its apparent persistence for small temperatures shown in~\cite{amico1}. Considering the thermal behavior of the correlations for the finite case of $N=5$ spins, one finds the same qualitative features of the factorization field as in the thermodynamic limit, {\it i.e.} a constant value of QD among all pairs of spins when $\lambda=\lambda_f$, which persists for remarkably large values of $T$ (up to $\sim0.1$ for $\gamma=0.5$). Examining the energy spectrum shown in Fig.~\ref{energylevels} {\bf (d)} we see a large gap between the first and second excited energy levels, with the exact difference being dependent on $\gamma$. This means that a significant amount of thermal energy is required before the higher order states become occupied. When $\gamma=0.5$, for  $0<T<0.1$ the thermal energy is insufficient to do this and the ground and first excited states are the only ones occupied with their degeneracy point remaining at $\lambda=\lambda_f$. However when $T$ is large enough to excite the higher energy levels a separability region appears. This explanation can also be confirmed for small chains by determining the rank of the full thermal density matrix 
\begin{equation}
\varrho(T)=\frac{e^{-\hat{\mathcal{H}}/{T}}}{\mathrm{Tr}[e^{-{\hat{\cal H}}/{T}}]}.
\end{equation}
For $N=5$, $\gamma=0.5$ and $T<0.1$ the rank is two, confirming that the thermal state is a mixture of the ground and first excited states, and they become equally mixed at $\lambda=\lambda_f$. The rank increases when $T>0.13$ and higher order states have become occupied. As we increase $N$ the gap between first and second excited energy levels reduces, which is why, in the thermodynamic limit, factorization only exists for relatively small values of $T\!<\!\!0.03$, as discussed in~\cite{amico1}. 

\begin{figure}[t]
{\bf (a)}\hskip3.5cm{\bf (b)}\\
\includegraphics[scale=0.4]{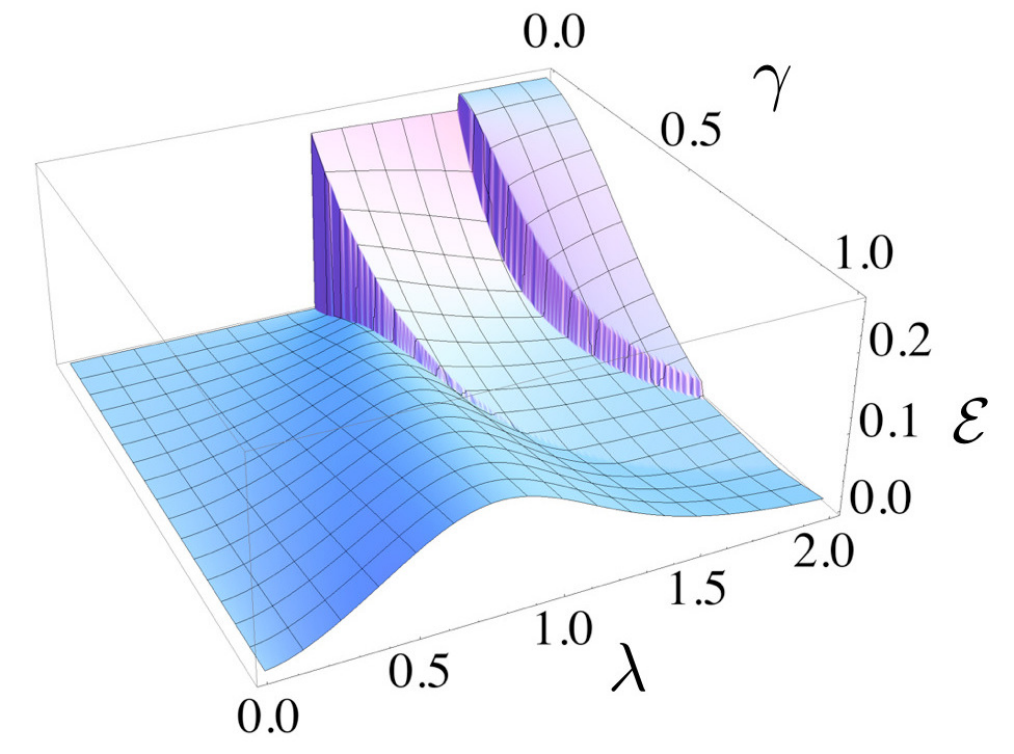}~~\includegraphics[scale=0.4]{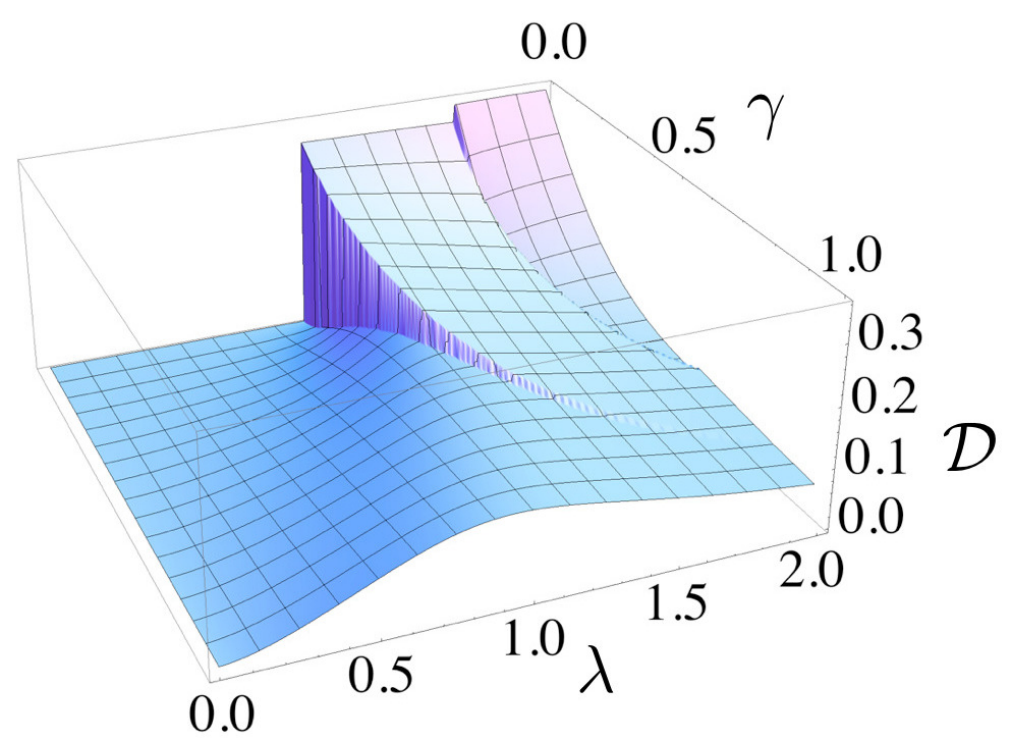}\\
{\bf (c)}\hskip3.5cm{\bf (d)}\\
\includegraphics[scale=0.4]{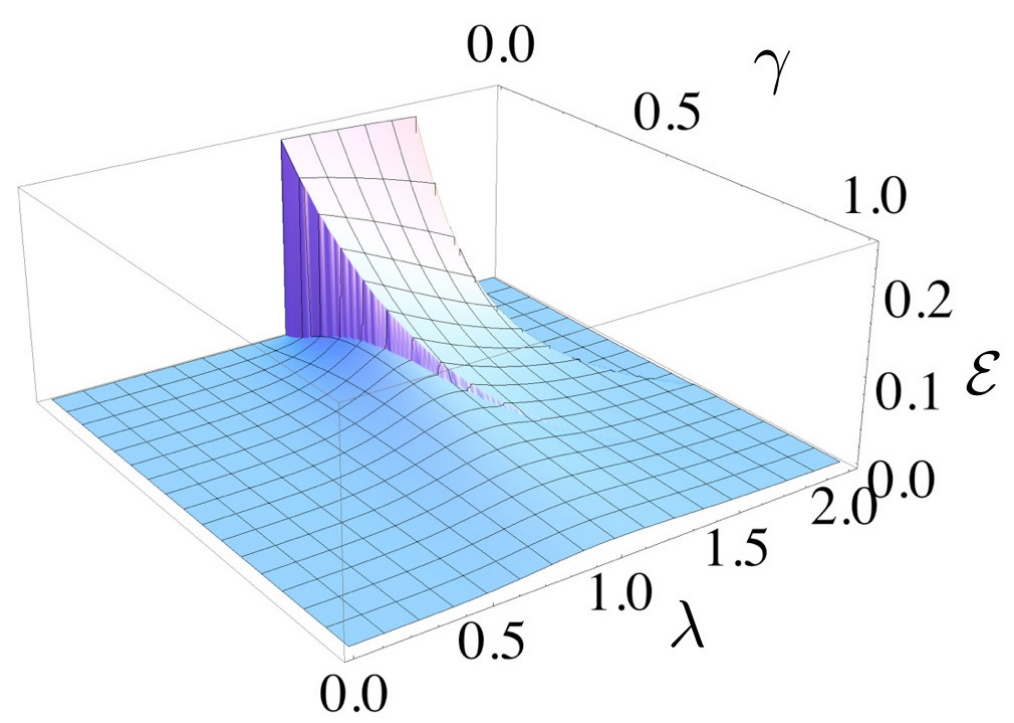}~\includegraphics[scale=0.4]{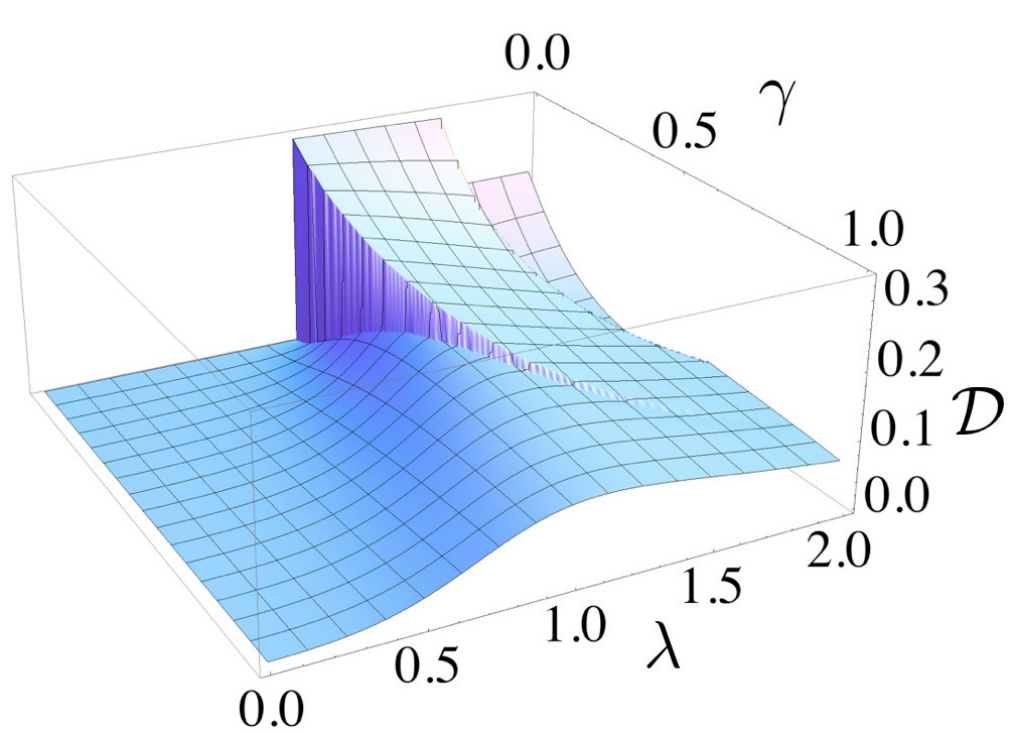}
\caption{Finite size, $N=5$, correlations in the reduced states of nearest neighbor pairs through {\bf (a)} EoF and {\bf (b)} QD, and next-nearest neighbor pairs {\bf (c)} EoF and {\bf (d)} QD.}
\label{finite5qubits}
\end{figure}

\begin{figure}[t]
{\bf (a)}\hskip3.5cm{\bf (b)}\\
\includegraphics[scale=0.4]{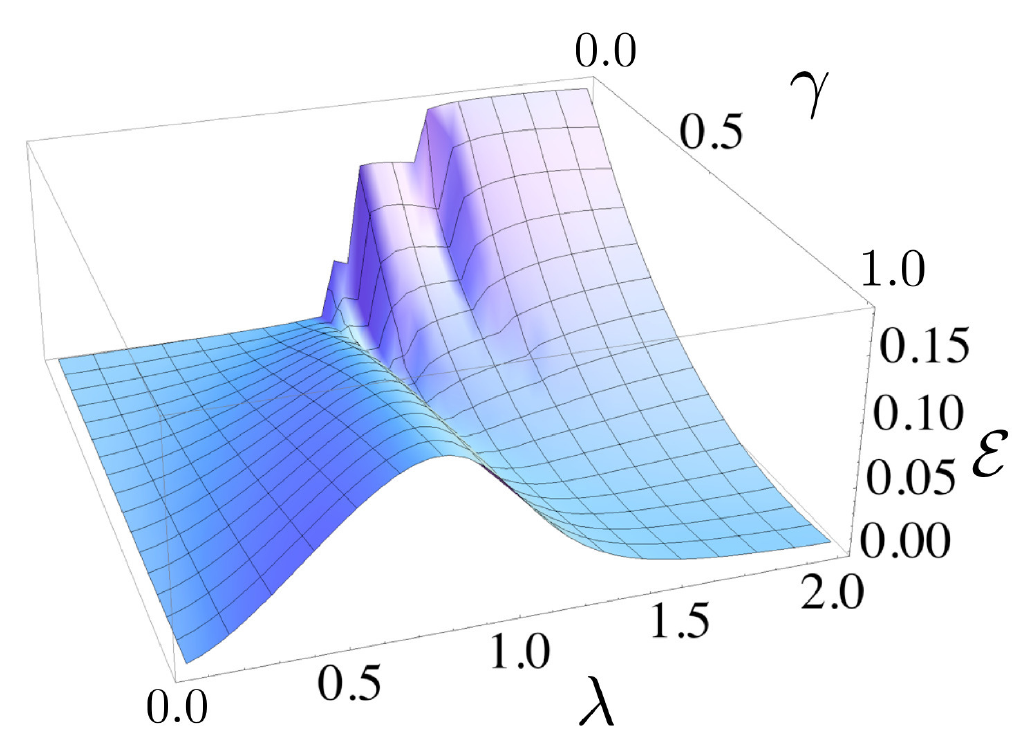}~~\includegraphics[scale=0.4]{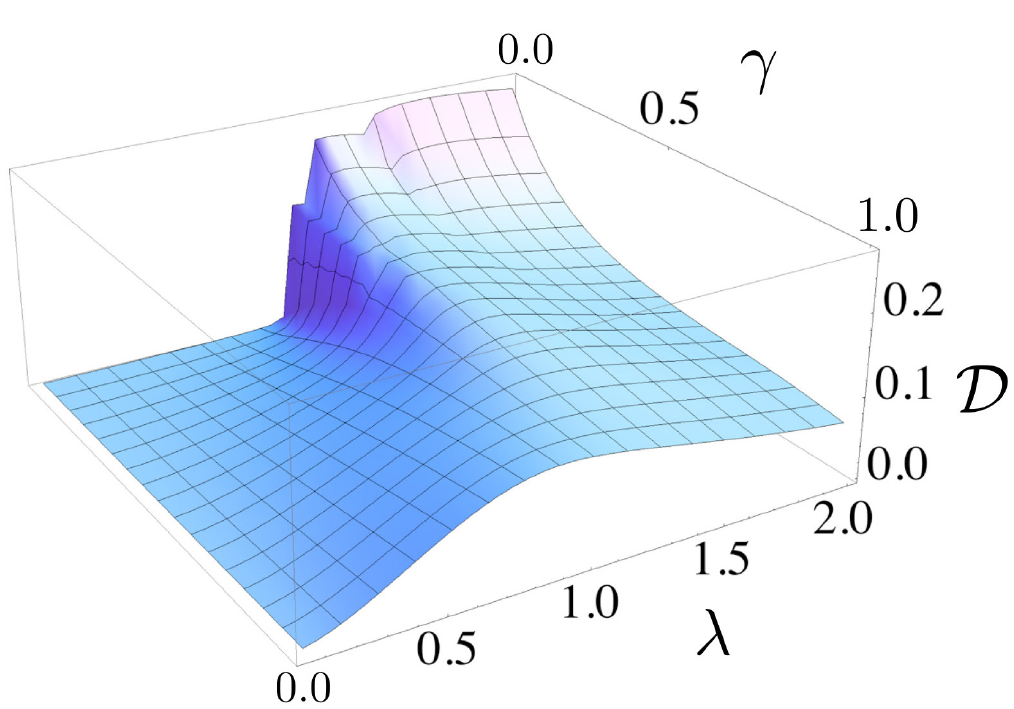}\\
{\bf (c)}\hskip3.5cm{\bf (d)}\\
\includegraphics[scale=0.4]{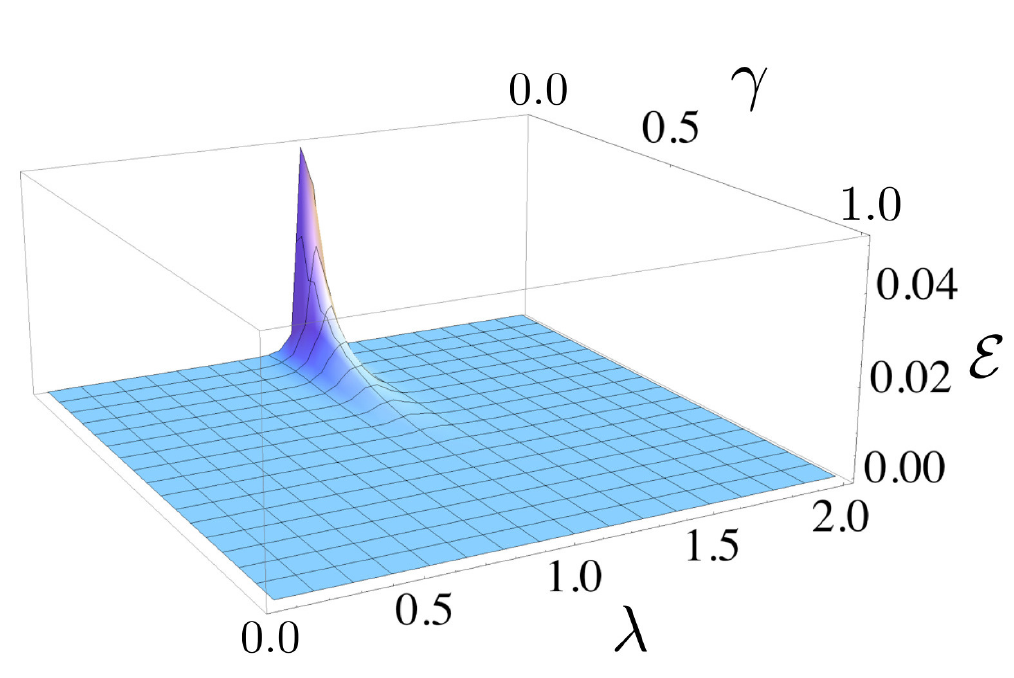}~~\includegraphics[scale=0.4]{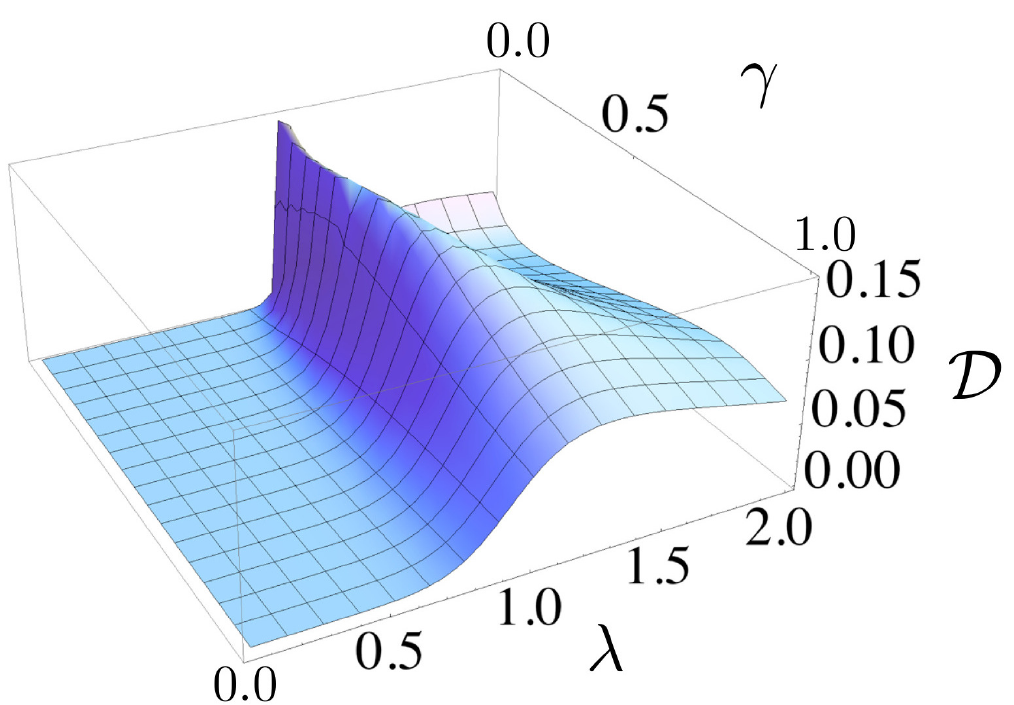}
\caption{Finite size, $N=10$, correlations in the reduced states of nearest neighbor pairs {\bf (a)} EoF and {\bf (b)} QD. (Quasi-)Long-range [$r$=5] {\bf (c)} EoF and {\bf (d)} QD.}
\label{finite10qubits}
\end{figure}

For $T\!\to\!0$ the different parities of the ground and first excited state leads to each exhibiting a different type of bipartite entanglement. For $\lambda<\lambda_f$ it is parallel entanglement, while for $\lambda>\lambda_f$ it is antiparallel~\cite{palma}. When $\lambda=\lambda_f$ and for small $N$ the entanglement of nearest neighbor spin pairs in either of the degenerate ground states is non-zero, but expressing the ground state as a mixture of these degenerate states leads to an overall decrease in the bipartite entanglement shared among the spins. As the states are highly symmetric and due to the constraints on the shareability of entanglement, larger $N$ results in smaller bipartite entanglement in both states, and with the mixing decreasing it further it quickly approaches zero for $N\gtrsim10$. Understanding that factorization is due to this energy level crossing also explains succinctly why there is no such phenomena in the extremal case of $\gamma=1$, {\it i.e.} the Ising model. In this instance the energy spectrum is always non-degenerate for any arbitrary non-zero magnetic field. Therefore, as there is no energy level crossing, there is no ground state factorization. 

The features of the finite-size thermal ground state are clearly seen by examining Figs.~\ref{finite5qubits} and \ref{finite10qubits}. In Fig.~\ref{finite5qubits} {\bf (a)} and {\bf (b)} we show the nearest neighbor EoF and QD for the case of $N=5$. The sudden changes in the correlations in the ground state correspond to the same parameters as for the energy level crossings and we see the same qualitative behavior for next-nearest-neighbors shown in panels {\bf (c)} and {\bf (d)}. The small size of the chain means that the short range nature of the interaction is significant among all parties, making identifying signatures of the critical nature of the model difficult to be witnessed by simply examining pairs of spins. However it should be noted that by employing global measures one can capture the critical nature as shown in~\cite{campbellGD}. Figure~\ref{finite10qubits} shows the nearest-neighbor EoF [panel {\bf (a)}] and QD [panel {\bf (b)}] and the quasi long-range, {\it i.e.} spins separated by 5 sites, EoF [panel {\bf (c)}] and QD [panel {\bf (d)}] for $N=10$. Already for such a small chain we can see more clearly the qualitative features of the model. Interestingly, here the quasi long-range behavior is very similar to the thermodynamic limit, the EoF is almost universally zero while QD shows a sharp change in the vicinity of the critical point, panels {\bf (c)} and {\bf (d)}. 

\begin{figure}[t]
{\bf (a)}\\
\includegraphics[scale=0.7]{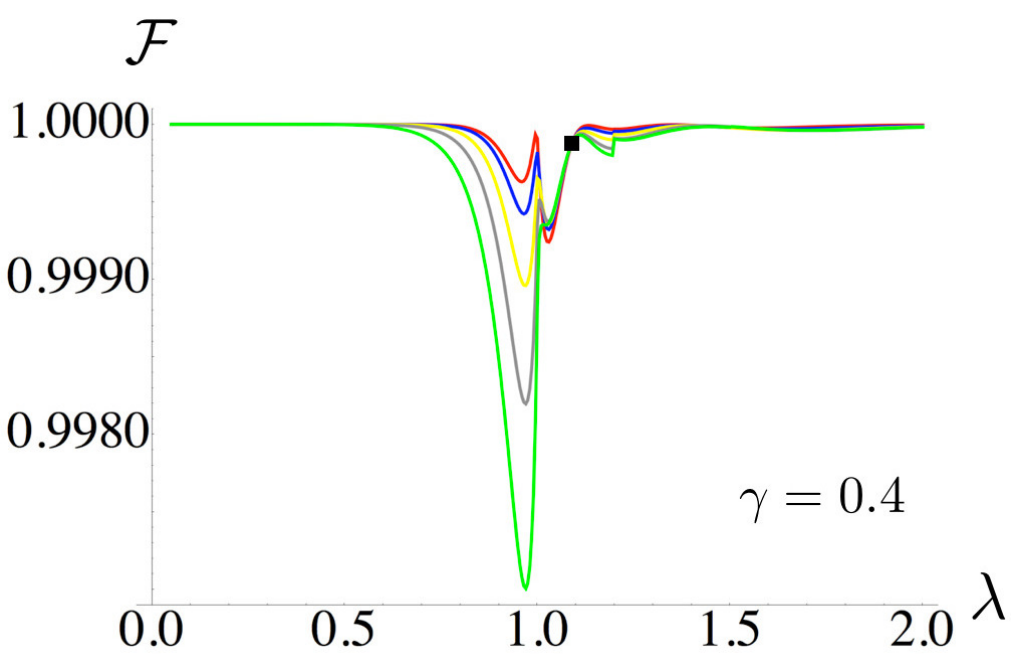}\\
{\bf (b)}\\
\includegraphics[scale=0.7]{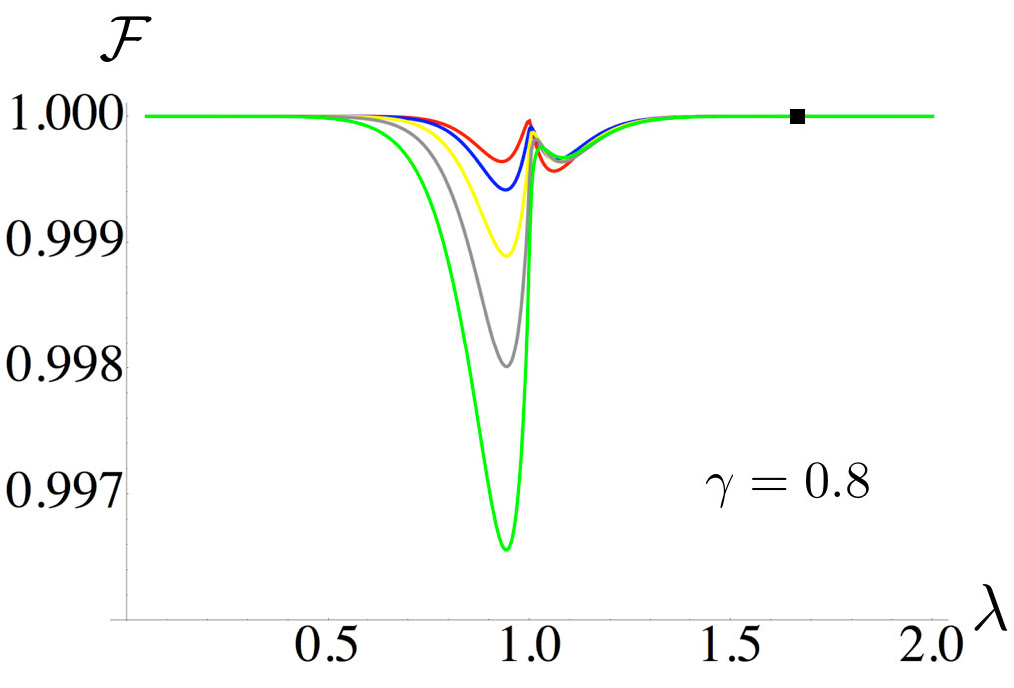}
\caption{Fidelity, $\mathcal{F}$, of the reduced state calculated from a finite-size chain with $N=10$ with the full thermodynamic reduced state for {\bf (a)} $\gamma=0.4$ and {\bf (b)} $\gamma=0.8$. In all panels, top-most curves correspond to nearest-neighbor pairs, and each subsequent descending curve corresponds to an increase in site separation of 1.}
\label{10qubitsFid}
\end{figure}

\section{Finite-size State Fidelity}
\label{fidelity}
Finally we address the question of how similar finite sized systems are to the thermodynamic limit by calculating the fidelity between the states arising from both situations as 
\begin{equation}
\mathcal{F}(\sigma,\rho)=\text{Tr}\left[ \sqrt{ \sqrt{\sigma} \rho \sqrt{\sigma} } \right].
\end{equation}
In Fig.~\ref{10qubitsFid} we show the fidelity for all possible bipartite reduced states for $N=10$ with the corresponding state arising from the solution in the thermodynamic limit with the same separation between spins in the chain for $\gamma=0.4$ and $\gamma=0.8$. Strikingly one can see that already for this small number of spins the states are almost identical and we find $\mathcal{F}>0.995$ for all separations and both anisotropies. While such a high fidelity is in part due to the periodic boundary conditions, given the disparity between the system sizes it is still a remarkably large value. In fact, the fidelity only changes in the vicinity of the critical point and for nearest neighbors the effect is extremely small with $\mathcal{F}>0.9995$. As the separation is increased the effect becomes more pronounced, however a fidelity of $\mathcal{F}>0.995$ is still maintained. In each panel the position of the factorization point, $\lambda_f$, is represented by a black square, and at these points we find $\mathcal{F}>0.9999$ between the two states independent of $r$. This means that at the factorization point the infinite and finite states are virtually identical. While the small change in the fidelity in the vicinity of the critical point indicates that the QPT is only truly manifest in the thermodynamic limit, strong evidence can nevertheless be found by examining small chains of $N=10$. The fact that $\mathcal{F}>0.9999$ at the factorization point gives further evidence that this phenomena is due to the mixing of highly symmetric energy levels of different parity which is a property independent of particle number. These observations together show that the QPT and ground-state factorization are manifestly different phenomena.

\section{Conclusions}
\label{conclusions}
We have presented an analysis of the long-range thermal quantum correlations in the anisotropic $XY$-model. Our results indicate that the long-range quantum discord is a versatile tool with which to study criticality. While the nearest-neighbor interaction strongly affects the properties at short ranges, at a suitably large site separation global features become more important and the quantum discord allows to faithfully capture the QPT and reflects the qualitative features associated with the QPT mechanism. The advantages of the long-range QD have been shown to not be restricted to zero temperature and we have found that it is possible to estimate the critical point for finite $T$ from a simple function. By considering small finite size systems we have shown the factorization phenomenon in this model can be fully explained in terms of the systems spectrum, and therefore that ground-state factorization and {\it bonafide} QPTs are manifestly different phenomena. 

\acknowledgments
JR gratefully acknowledges the hospitality of OIST Graduate University through the research intern scheme where the early stages of this work were performed and financial support from the EPSRC (UK). N.L.G. gratefully acknowledges the hospitality of OIST Graduate University and funding from the MIUR under the FIRB 2012 RBFR12NLNA. The authors gratefully acknowledge useful discussions with Gerardo Adesso, Luigi Amico, Gian Luca Giorgi, Mauro Paternostro, and Tommaso Tufarelli.

\vskip0.5cm
{\it Note added:} On completion of this work we became aware of two related papers. In Ref.~\cite{hofmann} scaling of genuine multipartite entanglement in a quantum phase transition is shown for the same model. While in Ref.~\cite{giampaolo} the authors address some complementary questions to those studied here through the use of multipartite entanglement.


\begin{thebibliography}{99}
\bibitem{energynote} We show up to $N=5$ only for clarity of figures. the energy level crossing is there for all $N$ as shown in~\cite{gianluca}.
\end{thebibliography}
\end{document}